\documentclass[aps,pra,twocolumn,superscriptaddress]{revtex4-1}

\usepackage{graphicx}
\usepackage{hyperref}
\usepackage{amsmath}
\usepackage{amssymb}
\usepackage{epstopdf}
\usepackage{multirow}
\usepackage{threeparttable}
\usepackage[usenames]{color}

\begin{document}

\title{Mid-infrared Fourier ptychographic upconversion imaging}

\author{Tingting Zheng}
\thanks{These authors contributed equally to this work.}
\affiliation{State Key Laboratory of Precision Spectroscopy, East China Normal University, Shanghai 200062, China}

\author{Zhuohang Wei}
\thanks{These authors contributed equally to this work.}
\affiliation{State Key Laboratory of Precision Spectroscopy, East China Normal University, Shanghai 200062, China}

\author{Kun Huang}
\email{khuang@lps.ecnu.edu.cn}
\affiliation{State Key Laboratory of Precision Spectroscopy, East China Normal University, Shanghai 200062, China}
\affiliation{Chongqing Key Laboratory of Precision Optics, Chongqing Institute of East China Normal University, Chongqing 401121, China}
\affiliation{Collaborative Innovation Center of Extreme Optics, Shanxi University, Taiyuan, Shanxi 030006, China}

\author{Mengyao Yu}
\affiliation{State Key Laboratory of Precision Spectroscopy, East China Normal University, Shanghai 200062, China}

\author{Jianan Fang}
\affiliation{State Key Laboratory of Precision Spectroscopy, East China Normal University, Shanghai 200062, China}

\author{Zhaoyang Wen}
\affiliation{State Key Laboratory of Precision Spectroscopy, East China Normal University, Shanghai 200062, China}

\author{Jixi Zhang}
\affiliation{State Key Laboratory of Precision Spectroscopy, East China Normal University, Shanghai 200062, China}

\author{Heping Zeng}
\email{hpzeng@phy.ecnu.edu.cn}
\affiliation{State Key Laboratory of Precision Spectroscopy, East China Normal University, Shanghai 200062, China}
\affiliation{Chongqing Key Laboratory of Precision Optics, Chongqing Institute of East China Normal University, Chongqing 401121, China}
\affiliation{Shanghai Research Center for Quantum Sciences, Shanghai 201315, China}
\affiliation{Chongqing Institute for Brain and Intelligence, Guangyang Bay Laboratory, Chongqing, 400064, China}

\begin{abstract}
Frequency upconversion technique offers an appealing approach for sensitive mid-infrared (MIR) imaging at room temperature. However, the spatial resolution of the upconversion imager has been notoriously restricted by the limited transverse section of the involved nonlinear crystal at the Fourier plane. Here, we implement a wide-field and high-resolution MIR upconversion imaging system based on elliptical pumping and Fourier ptychography. Specifically, an elliptical pump beam is engineered to accommodate the narrow aperture of chirped-poling crystals, thus facilitating the acquisition of high spatial frequency components in the lateral direction. Such an elliptical passband in the Fourier space is then discretely rotated to generate a sequence of upconversion images, which allows computational recovery of a high-resolution object image through a combination of synthetic aperture and phase retrieval operations. Consequently, an enhanced spatial resolution of 39 $\mu$m is achieved within a field of view about 25 mm, which corresponds to a space-bandwidth product of 3.2$\times$10$^5$, over tenfold larger than previously demonstrated values. Moreover, the MIR upconversion imager can operate under a low-light illumination of 1 photon/pulse/pixel. Therefore, the presented paradigm of nonlinear Fourier ptychography paves the way toward high-throughput infrared imaging with massive resolvable elements and single-photon sensitivity, which would stimulate a variety of applications such as industry inspection and biomedical diagnosis.
\end{abstract}

\maketitle

\section{Introduction}
Mid-infrared (MIR) imaging provides unique information about the thermal and molecular characteristics of objects \cite{Vodopyanov2020Book}, which has become a powerful tool in a wide range of scientific, industrial, and military applications, such as industrial inspection, medical diagnostics, environmental monitoring, and defense security \cite{Cheng350Science, Hermes2018JO, Lau2023CSR}. Moreover, highly sensitive MIR imaging is particularly demanded in many extreme scenarios, for instance, remote sensing at a long stand-off distance \cite{Hadfield2023Optica}, deep penetration through thick materials \cite{Israelsen2019LSA}, phototoxicity-free examination of biological samples \cite{Shi2020NM}, and non-destructive analysis of heritage artifacts \cite{Wolley2023PNAS}. However, current focal plane arrays (FPAs) typically suffer from limited detection sensitivity due to severe dark current and thermal noise in narrow-bandgap semiconductor devices, especially at the room-temperature operation \cite{Razeghi2014RPP, Wang2019Small}. Notably, emerging MIR sensors based on low-dimensional materials have shown great potential in elevating the operation temperature, albeit with an attainable noise equivalent power far away from the single-photon level \cite{Wu2021NR}. Recently, superconducting nanowire detectors under the cryogenic condition have demonstrated excellent photon-counting capability across a wide spectral coverage into the far-infrared region \cite{Taylor2023Optica, Chen2021SB, Pan2022OE}, yet large format arrays with massive pixels are still in the infancy \cite{Hampel2024APL}. Therefore, it calls for continuous endeavors to develop room-temperature MIR single-photon imagers with high spatial resolution and wide field of view \cite{Tamamitsu2024NP, Hiramatsu2019SA, Pavlovetc2020PCCP, Park2021AP}.

Alternatively, frequency upconversion strategy provides an indirect yet effective solution to address the aforementioned challenges \cite{Barh2019AOP}. The involved nonlinear conversion process enables high-performance detection of infrared light with more mature visible or near-infrared detectors, thereby expanding the capabilities and accessibility of infrared imaging technologies \cite{Dam2012NP, Wang2021LPR, Mrejen2020LPR, Huot2019AO, Zeng2023LPR, Paterova2020SA, Kviatkovsky2020SA, Rehain2020NC, Ge2023PRAppl}. Nowadays, the upconversion architecture has been widely adopted to implement various ultra-sensitive MIR imaging modalities with desirable features of single-photon detection sensitivity \cite{Dam2012NP, Wang2021LPR}, MHz-level frame rate \cite{Huang2022NC}, high-precision timing stamp \cite{Fang2023LSA}, and hyperspectral resolving capability \cite{Zhao2023NC, Fang2024NC, Junaid2019Optica}. In these demonstrations, second-order bulky crystals are usually used to access high nonlinear coefficients and long interaction lengths \cite{Dam2012NP, Wang2021LPR, Mrejen2020LPR, Huot2019AO}, comparing to other platforms based on third-order two-photon absorption \cite{Knez2022SA} or ultra-thin metasurfaces \cite{Molina2024AM}. Furthermore, the use of periodical poling nonlinear crystals allows one to leverage the quasi-phase-matching (QPM) condition to achieve a high conversion efficiency with a modest pump intensity \cite{Bahabad2010NP}. However, the thickness of periodically poled crystals is typically at the millimeter scale, which severely limits the passband aperture at the Fourier plane of an upconversion imaging system in the 4f configuration \cite{Barh2019AOP}. 

Indeed, periodic poling of large aperture crystals poses a technical challenge, which requires very high voltages about tens of kV/mm to access a sufficiently strong coercive electric field during the polarization switching process \cite{Missey1998OL}. Moreover, the obtained periodic domain structures inevitably suffer from degraded homogeneity due to duty cycle variations, domain broadening and domain merging throughout the crystal thickness. Although attempts at fabricating large-aperture crystals have been reported for long QPM periods \cite{Ishizuki2012OE, Ishizuki2005OL}, yet small poling domains below 20 $\mu$m remain inaccessible. Notably, diffusion bonding a stack of crystal plates offers an intriguing way to increase the effective aperture size \cite{Missey1998OL}, yet with remaining concerns on grating alignment and operation robustness. To date, it is imperative to develop novel techniques for circumventing the bottleneck of limited Fourier apertures, thus improving the spatial resolution of the upconversion imagers to promote widespread applications in practice.

In this work, we address the long-sought-after quest by resorting to the concept of Fourier aperture synthesis, which enables computational recovery of an image with much higher resolution than what a single aperture could achieve \cite{Huang2024LSA}. Among others, Fourier ptychography microscopy has been evolved as one of the most representative modalities in the realm of synthetic aperture imaging \cite{Pan2020RPP, Konda2020OE, Zheng2021NRP}. The Fourier ptychographic imaging retrieves the phase information from a set of acquired intensity patterns with an iterative reconstruction algorithm, thereby eliminating the interferometry complexity of direct phase measurements, as exist in holographic approaches \cite{Zheng2021NRP}. Accordingly, the technique does not require a reference beam or much additional optical hardware, which makes it particularly well-suited for applications within standard imaging systems \cite{Zheng2013NP, Holloway2017SA}. 

Here, we extend the design strategy of the Fourier ptychography into the nonlinear upconversion imaging architecture. Specifically, an elliptical pump is used to adapt the narrow aperture of the crystal plate, and a subsequent rotational sampling in the Fourier space results in an enlarged effective aperture size defined by the crystal width. In contrast to the thickness, the other dimensions of the crystal can be quite large, currently limited only by the size of the standard wafers. In our proof-of-principle experiment, the spatial resolution of the MIR upconversion imager is enhanced at least three times than that for the conventional Gaussian pumping in the presence of a nonlinear crystal with a transverse section of 1 mm (thickness) $\times$ 3 mm (width). Moreover, the chirped poling design of the nonlinear crystal favors a large field of view about 25 mm, which results in an unprecedentedly large number of resolvable elements up to 3.2$\times$10$^5$. In parallel, the coincidence pulsed pumping facilitates a high-efficiency and low-noise conversion process, which permits an ultra-sensitive MIR imaging under a low-light illumination flux of 1 photon/pulse/pixel. Therefore, the presented MIR Fourier ptychographic upconversion imaging system features single-photon sensitivity, high spatial resolution, and wide field of view, which would provide an enabling tool in high-throughput examination of chemical and biomedical materials.

\section{Basic principles}
\begin{figure*}[t!]
\includegraphics[width=0.66\textwidth]{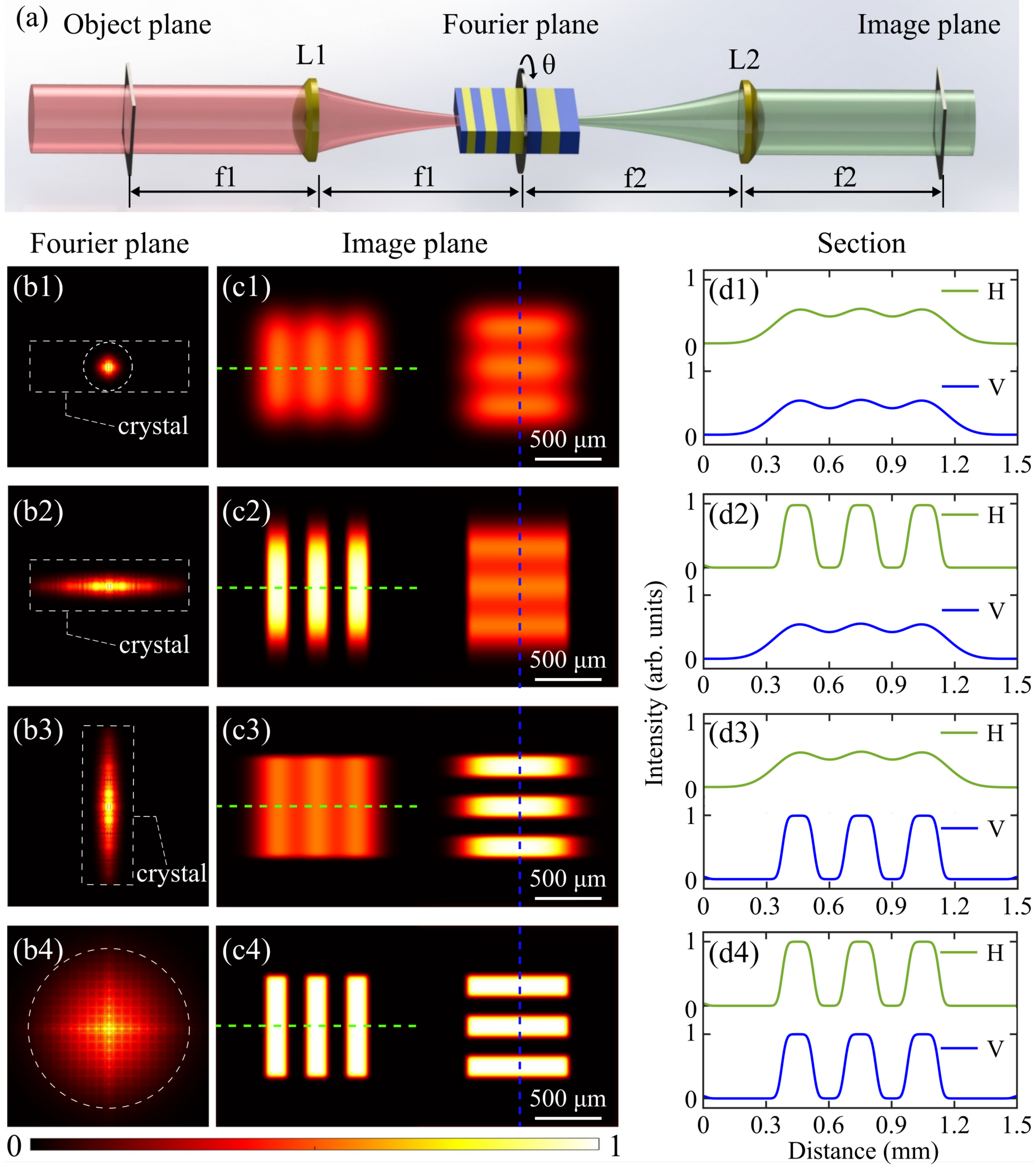}
\caption{Conceptual illustration for MIR Fourier ptychographic upconversion imaging. (a) Diagram of a typical MIR upconversion imaging system in the 4f configuration, where a nonlinear crystal is located at the Fourier plane as a low-pass filter for spatial frequencies. (b1-b4) Four pumping schemes at the Fourier plane with a Gaussian beam (b1), a horizontal elliptical beam (b2), a vertical elliptical beam (b3), and a stitched aperture with rotated elliptical beams (b4). The diameters of the dashed circles in (b1) and (b4) correspond to the lengths of short and long axes for the elliptical beam, respectively. (c1-c4) Numerically calculated upconversion images, which indicate a substantial improvement on the spatial resolution through the aperture synthesizing operation. (d1-d4) Corresponding cross sections for the object bars along the horizontal (H) and vertical (V) directions, as denoted by the dashed lines in the images (c1-c4).}
\label{fig1}
\end{figure*}

Figure \ref{fig1}(a) presents a typical upconversion imaging system based on a 4f configuration, where the nonlinear crystal is located at the Fourier plane. The involved conversion process relies on sum-frequency generation (SFG) under the non-depleted pump condition. The wavelengths for the signal, pump, and upconverted optical fields satisfy the energy conservation law as $1/\lambda_\text{up} = 1/\lambda_\text{s} + 1/\lambda_\text{p}$. In the presence of an object with an intensity distribution $I_\text{object}(x, y)$, the upconverted image, under the conditions of incoherent illumination and perfect phase matching, can be expressed as \cite{Barh2019AOP}:
\begin{equation}
I_\text{up}(x, y) = I_\text{object}(-\frac{\lambda_s f_1}{\lambda_\text{up} f_2} x, -\frac{\lambda_s f_1}{\lambda_\text{up} f_2} y) \ast  \text{PSF} \ , 
\label{eq1}
\end{equation}          
where $f_{1,2}$ are the focal lengths of the lenses in the 4f imaging system, the asterisk symbol denotes the two-dimensional convolution operation, and PSF is the point spread function. It can be seen that the object field is transferred to the image plane with a magnification of $M = - (\lambda_\text{up} f_2) / (\lambda_s f_1)$. Moreover, the resulting image is blurred by the aperture effect at the Fourier plane, which is ascribed to the beam size of the pump or the truncation size of the crystal, whichever is smaller. For a Gaussian pump with a beam radius of $\omega_p$ (defined as 1/$e^2$ of the maximum intensity), the incoherent PSF is given as:
\begin{equation}
\text{PSF}_\text{Gaussian} = \frac{2 \pi \omega_p^2}{\lambda_\text{up}^2 f_2^2}\exp[-\frac{2 \pi^2 \omega_p^2(x^2+y^2)}{\lambda_\text{up}^2 f_2^2}] \ ,
\label{eq2}
\end{equation} 
which is formed by the Fourier transform of the Gaussian pump field profile, evaluated at spatial frequencies at $x/(\lambda_\text{up} f_2)$ and $y/(\lambda_\text{up} f_2)$. Consequently, the spatial resolution at the object plane is inferred to be \cite{Junaid2019Optica}
\begin{equation}
\mathcal{R} =  \frac{\sqrt{2} f_1 \lambda_s}{\pi \omega_p} \ .
\label{eq3}
\end{equation} 

In practice, the pump beam is usually designed to be smaller than the transverse size of the crystal. The soft Gaussian aperture is ultimately limited by the crystal thickness. Nowadays, periodically poled crystals with a thickness up to 1 mm can be readily obtained with good reproducibility, making this the standard thickness for commercially available QPM crystals. The underlying difficulty in fabricating large-aperture crystals with short poling periods is mainly due to the stringent demand on the crystal homogeneity to ensure that the inverted domains propagate uniformly throughout the device thickness during the poling process  \cite{Missey1998OL}. Figures \ref{fig1}(b1-d1) show the simulated imaging performance under the conventional Gaussian pumping. The crystal aperture is set to be 1 mm (thickness) $\times$ 3 mm (width), and the bar width of the target is 150 $\mu$m. In this case, horizontal and vertical bars can not be clearly resolved due to stringent low-pass filtering for the spatial frequency components in the Fourier domain, as indicated from the cross sections in Fig. \ref{fig1}(d1).

\begin{figure*}[t!]
\includegraphics[width=0.84\textwidth]{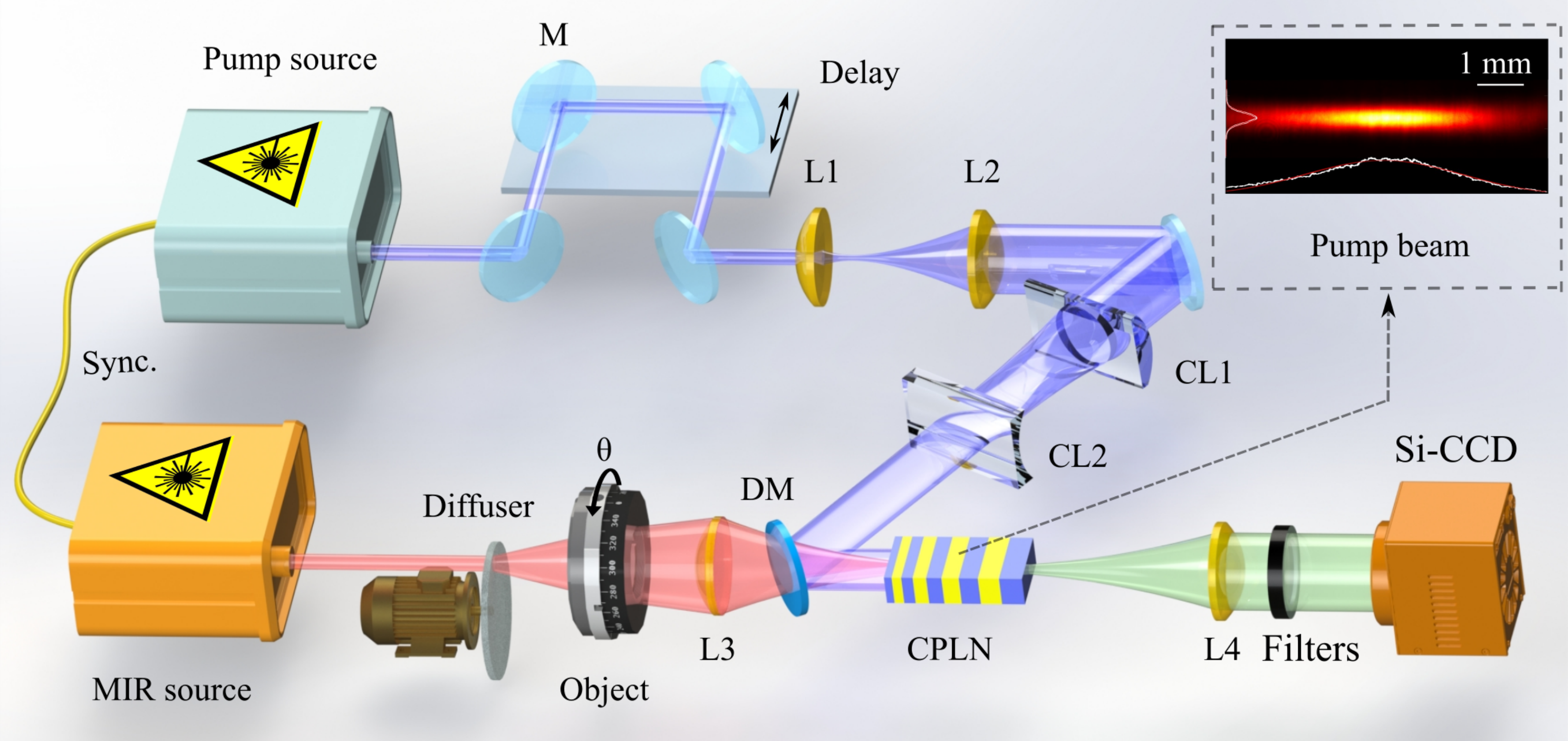}
\caption{Experimental setup for the MIR upconversion imaging based on Fourier ptychography. The pump and signal sources are from two temporally synchronized pulsed lasers at 1030 nm and 3070 nm, respectively. The MIR signal beam passes through a rapidly spinning diffuser to emulate an incoherent illumination, which covers a transmission object with a diameter about 1 inch. The transmitted light is collected by a calcium fluoride lens (L3), and then focuses into a nonlinear crystal based on chirped-poling lithium niobate (CPLN). Meanwhile, the pump beam is enlarged by a beam expander via a pair of lenses (L1 and L2). The expanded beam is shaped to form an elliptical beam as illustrated in inset by using a pair of cylindrical lenses (CL1 and CL2). The elliptical pump is engineered to adapt the crystal aperture for addressing more spatial-frequency components as much as possible. The pump and signal beam are spatially mixed by a dichroic mirror (DM) into the CPLN crystal to perform the sum-frequency generation. The temporal overlap between the dual-color pulses is optimized to improve the conversion efficiency by tuning the delay line in the pump path. Finally, the upconverted image after a group of spectral filters is captured by a silicon-based charge coupled device (Si-CCD). Note that the object is mounted on a rotator. For each rotation angle $\theta$, the spatial frequency pattern in the Fourier domain will rotate correspondingly. Consequently, the elliptical bandpass effectively samples an enlarged Fourier space with a number of rotated apertures. The use of an iterative reconstruction algorithm based on alternating projection allows one to obtain a wide-field, high-resolution upconversion image, hence overcoming the limited numerical aperture dictated by the thin nonlinear crystal.}
\label{fig2}
\end{figure*}

As shown in Fig. \ref{fig1}(b2), an elliptical pump beam is preferable to accommodate the narrow aperture of the crystal, which favors interrogating more high-frequency components. The incoherent PSF is modified as
\begin{equation}
\text{PSF}_\text{elliptical} = \frac{2 \pi \alpha \omega_0^2}{\lambda_\text{up}^{2} f_2^2}\exp[-\frac{2 \pi^2 \omega_0^2( \alpha^2 x^2+y^2)}{\lambda_\text{up}^2 f_2^2}] \ ,
\label{eq4}
\end{equation} 
where $\alpha$ represents the ellipticity. In this case, the spatial resolution for the upconversion image in the horizontal direction is significantly enhanced, as shown in Figs. \ref{fig1}(c2) and (d2). Similarly, an orthogonal orientation of the elliptical pump along with a rotated crystal enables to enhance the vertical resolution, which is manifested in Figs. \ref{fig1}(b3-d3). We note that a conceptually equivalent operation is implemented in the experiment by simply rotating the object, which results in an effective rotation of the frequency pattern at the Fourier plane. 

Inspired by the spirit of aperture synthesis, we propose to adopt the Fourier ptychographic technique into the upconversion imaging, with an aim to alleviate the aperture limitation for approaching a high spatial resolution. To this end, we can record a series of upconversion images $\{I_\text{up}(x, y; \theta_k) \}$ at a predefined set of $N$ rotation angles $\{ \theta_k\}$ for the elliptical passband. Based on $N$ collected upconversion intensity images, we computationally reconstruct a high-resolution image of the object following a recovery procedure based on the alternating projection algorithm \cite{Zheng2021NRP, Zheng2013NP}. The adopted strategy is similar to that for the synthetic aperture imaging, which facilitates an expanded coverage in Fourier space via multi-image fusion as shown in Fig. \ref{fig1}(b4). Ideally, the permitted spatial resolution at the object plane can be modified as $\mathcal{R} =  \lambda_s f_1/ (\sqrt{2} D)$ under the limiting condition of a flat-top rectangular pumping, where $D$ is the width of nonlinear crystal \cite{Barh2019AOP}.

Notably, the Fourier ptychographic design requires no measured phase information, hence eliminating the system complexity associated with interferometric detection schemes. The reconstructed results are presented in Figs. \ref{fig1}(c4) and (d4), which show high-resolution performances along two orthogonal directions. The presented computational method for the MIR upconversion imaging based on Fourier ptychography is capable of providing a scalable spatial-bandwidth product (SBP) for existing upconversion imagers without involving setup modifications or phase measurements. More information on the numerical algorithm is given in Supplement 1, Note 1.

\begin{figure*}[t!]
\includegraphics[width=0.65\textwidth]{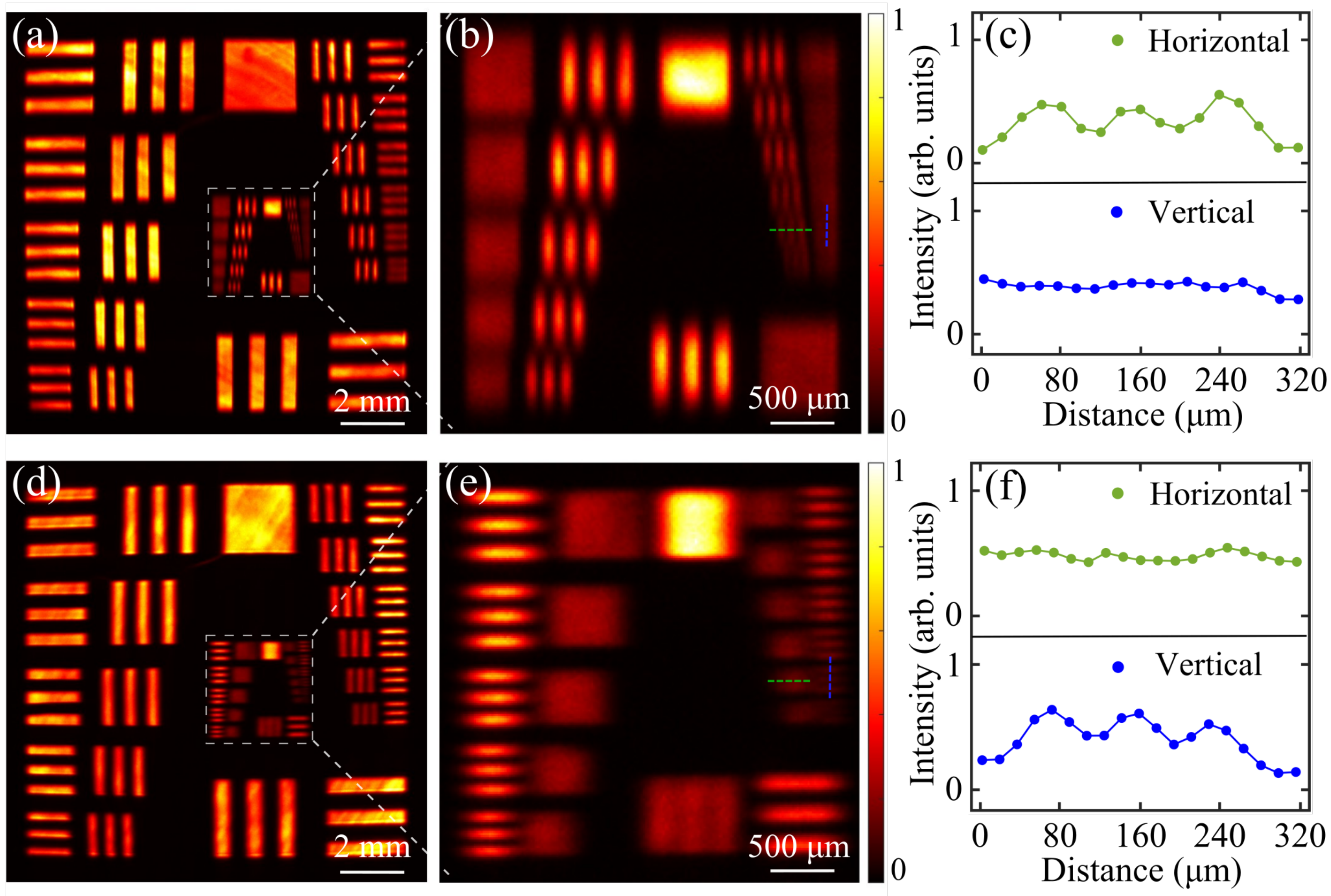}
\caption{Imaging performance characterization with a USAF resolution chart. (a, d) Upconversion images in the presence of elliptical pumping along the horizontal (a) and vertical (d) directions, respectively. (b, e) Central zones in dashed box are enlarged to show the high-resolution performance along the horizontal (b) and vertical (e) directions. The denoted bars in dashed lines are specified to be the fifth element in the third group, corresponding to a linewidth of 39 $\mu$m. (c, f) Corresponding cross sections, which indicate the selective resolution enhancement depending on the orientation of the elliptical pump.}
\label{fig3}
\end{figure*}

\section{Experimental Setup}
Figure \ref{fig2} illustrates the experimental setup of the MIR Fourier ptychographic imaging system. The whole configuration is based on the coincidence-pumping upconversion architecture, where the infrared signal at 3070 nm is spectrally converted by a train of temporally synchronized pump pulses at 1030 nm. The pulse durations of the signal and pump are 41.1 and 41.6 ps, respectively. The repetition rate is about 20 MHz. More details about the MIR light preparation and ultrashort pulse synchronization can be referred to our previous works \cite{Wang2021LPR, Fang2023LSA}. The MIR beam is expanded homogeneously via a rotating diffuser, which serves as a pseudo-thermal source for providing an incoherent illumination onto a transmission mask \cite{Junaid2019Optica, Fang2023LSA}. The diffused light from the transmitted pattern passes through a 4f imaging system based on two relay lenses L3 and L4 with focal lengths of 50 and 125 mm,  respectively. A chirped-poling lithium niobate (CPLN) is placed at the Fourier plane. The CPLN is fabricated with geometrical dimensions of 1 mm (thickness) $\times$ 3 mm (width) $\times$ 10 mm (length), and has linearly ramping poling periods from 19 to 24 $\mu$m along the propagation direction (see Supplement 1, Note 5). 

As for the pump source, the maximum output average power is about 2 W, which provides a peak power up to 2.4 kW. The Gaussian profile of the pump is enlarged via a beam expander before being shaped by two cylindrical lenses (CL). The one-dimensional focusing operation results in an elliptical beam as shown in the top-right corner in Figure \ref{fig2}. The full widths at $1/e^2$ along short and long axes are measured to be 0.5 mm and 5 mm, respectively. The corresponding peak intensity for the pump is estimated to be about 1.22 kW/mm$^2$, which allows for a maximum conversion efficiency of 0.1\%, comparable to previous works based on CPLN crystals \cite{Huang2022NC}. The elongated pump beam is designed to make full use of the transverse aperture of the crystal, which leads to a truncated beam with a better flatness along the horizontal direction. The pump and signal beams are spatially combined by a dichroic mirror (DM), before entering the CPLN crystal to perform the SFG. The temporal overlap between the two-color pulses is realized by tuning the delay line in the pump path, which helps to optimize the conversion efficiency \cite{Wang2021LPR, Huang2022NC}. The upconverted beam at 771 nm is filtered by a series of spectral filters to reject the parametric fluorescences and ambient noises \cite{Dam2012NP, Wang2021LPR}. The total transmittance of the filtering stage is 80\%, and the rejection ratio at the pump wavelength reaches to 220 dB. The filtered light is captured by a silicon-based electron-multiplying charge-coupled device (EMCCD, Andor iXonUltra 888). The involved low-noise conversion process and high-sensitivity photon detection are essential to realize the subsequent low-light-level MIR upconversion imaging \cite{Wang2021LPR}.

To implement the Fourier ptychographic operation, the object is fixed on a rotational mount to alter the spatial frequency pattern at the Fourier plane, which enables expansion of the Fourier passband following principle set forth with aperture synthesis. For a set of rotating angles, the captured upconversion images enable a high-resolution recovery of the object via a post-processing computation. The adopted iterative algorithm relies on the alternating projection procedure, which is realized by alternately constraining the amplitude to match the acquired low-resolution image sequence and the spectrum to match the rotating Fourier constraint \cite{Pan2020RPP}. See Note 1 in Supplement 1 for more details on the recovery process.

\begin{figure*}[t!]
\includegraphics[width=0.85 \textwidth]{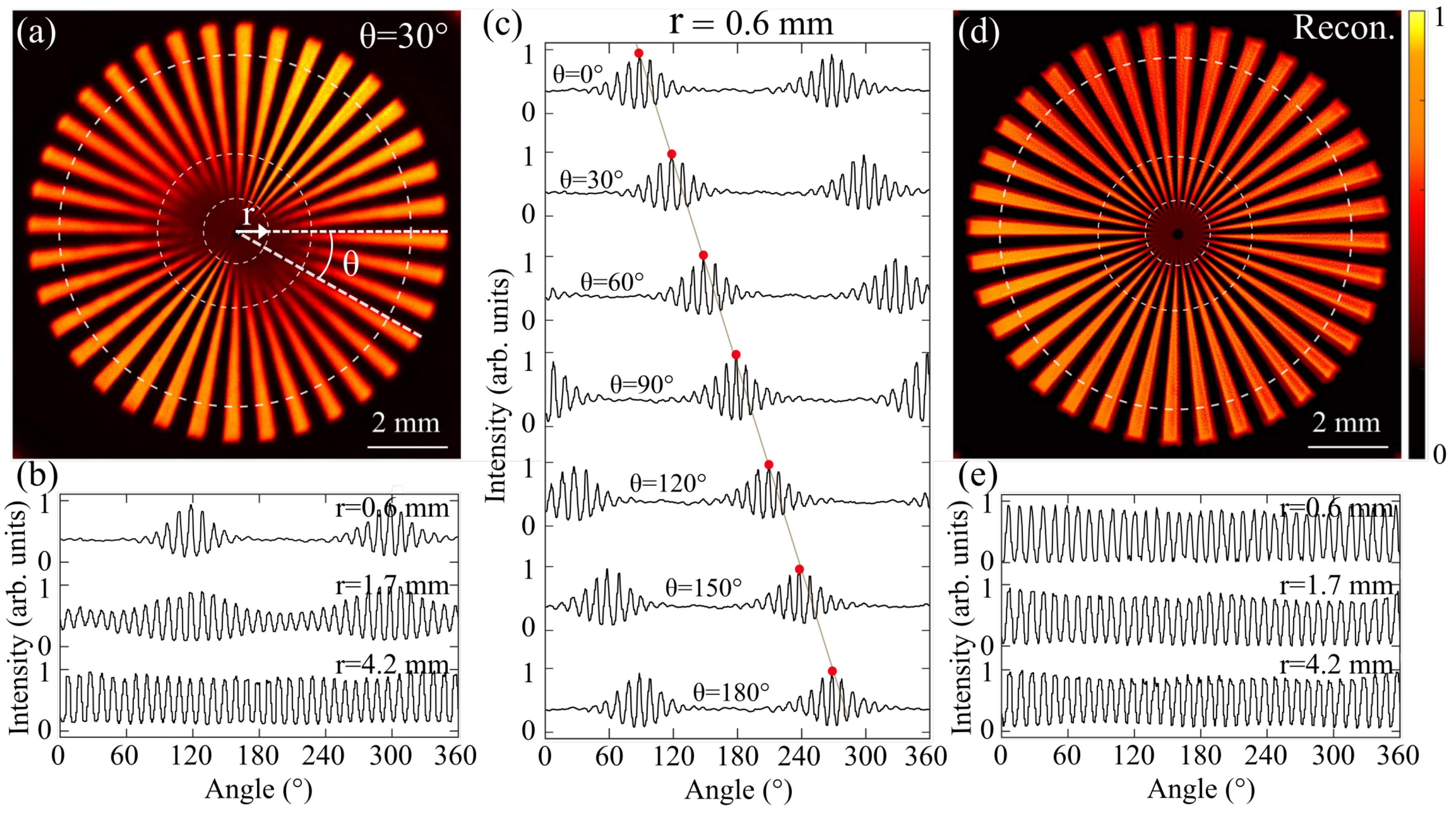}
\caption{High-resolution imaging reconstruction for a Siemens star resolution target. (a) Representative upconversion image under an elliptical pumping at $\theta = 30 ^\circ$. (b) Three circular cross sections at radii of $r$=0.6, 1.7, and 4.2 mm, which indicate a better ability for resolving elements at the angle of 120$^\circ$. (c) Recorded cross sections at $r$=0.6 mm in the case of various pumping angles $\theta$ from 0$^\circ$ to 180$^\circ$. The optimum resolving angle shows a linear dependence on the orientation angle of the pump. (d) Computationally recovered result after stitching the acquired sequence of upconversion images. (e) Corresponding cross sections at various radial distances, which demonstrate an enhanced spatial resolution at all angles.}
\label{fig4}
\end{figure*}

\section{Experimental results}
\subsection{Characterization of imaging performance}

Now we begin with characterizing the imaging performance with a USAF 1951 resolution target (Edmund, \#58-402). Figure \ref{fig3}(a) presents the captured upconversion image in the case of a horizontal elliptical pump. The central zone is zoomed in as shown in Fig. \ref{fig3}(b), which indicates a better resolution in the horizontal direction. The smallest resolvable bars denoted in dashed lines correspond to the fifth element in the third group, corresponding to a spatial linewidth of 39 $\mu$m. As a comparison, the vertically resolving limit is the fifth element in the first group, which corresponds to a resolution of 157 $\mu$m, as expected from the aperture limitation by the crystal thickness. The cross sections are presented in Fig. \ref{fig3}(c). The intensity contrast for the horizontal trace is measured to be 27\%, which is well beyond the value of 11\% at the Rayleigh criterion \cite{Barh2019AOP, Junaid2019Optica}. The achieved spatial resolution is close to the theoretical value of 36 $\mu$m. Similarly, the upconversion image exhibits resolution enhancement along the vertical direction in the case of vertical elliptical pumping, as shown in Figs. \ref{fig3}(d-f). Note that the field of view for the imaging system is about 25 mm, which is mainly limited by the one-inch optics used in the experiment \cite{Huang2022NC}. Accordingly, the number of the resolvable elements, or SBP, is calculated to be about 3.2$\times$10$^5$, as detailed in Supplement 1, Note 4. The obtained SBP is at least one order of magnitude more than previously reported results \cite{Huang2022NC, Fang2023LSA, Ge2023PRAppl, Junaid2019Optica}. The presented experimental observations agree t with the numerical simulations given in Supplement 1, Note 2.

\subsection{High-throughput MIR upconversion imaging}

Next, we turn to investigate the Fourier ptychographic upconversion imaging by using a Siemens star target (LBTEK, RTR1CH-N). The target has 36 sectors over 360$^\circ$. At the center of the target, there is a spot with a diameter of 200 $\mu$m, which favors precise alignment of the experimental setup to ensure the rotational axis of the object align with the principal axis of the imaging system. Moreover, the spot is also useful to determine the resolution of an optical system by assessing the proximity to the center of the pattern where an optical system is able to resolve adjacent bars. Figure \ref{fig4}(a) shows a representative upconversion image under an elliptical pumping at $\theta$=30$^\circ$. The circular cross sections at three radial distances $r$=0.6, 1.7, and 4.2 mm from the center are given in Fig. \ref{fig4}(b), which correspond to a bar width of 52, 148, and 367 $\mu$m, respectively. The intensity distribution as a function of the azimuthal angle indicates an orientation-selective resolution enhancement. 

Moreover, the angle-dependence enhancement is clearly manifested by the recorded cross sections along the same circle at $r$=0.6 mm in the case of various angles of the elliptical pumping, as shown in Fig. 4(c). At each pump angle $\theta$, the pattern with an optimal resolving performance is found to be located at $\theta+90^\circ$. In the experiment, we record a series of upconversion images at elliptical angles from 0$^\circ$ to 180$^\circ$ with an interval of 10$^\circ$. The acquired sequence of pupil segments are stitched in Fourier space via the Fourier ptychographic algorithm, which allows one to recover a high-resolution image in Fig. \ref{fig4}(d). The cross section at r = 0.6 mm in Fig. \ref{fig4}(e) exhibit a high resolution along all the directions.

\begin{figure}[b!]
\includegraphics[width=0.9\columnwidth]{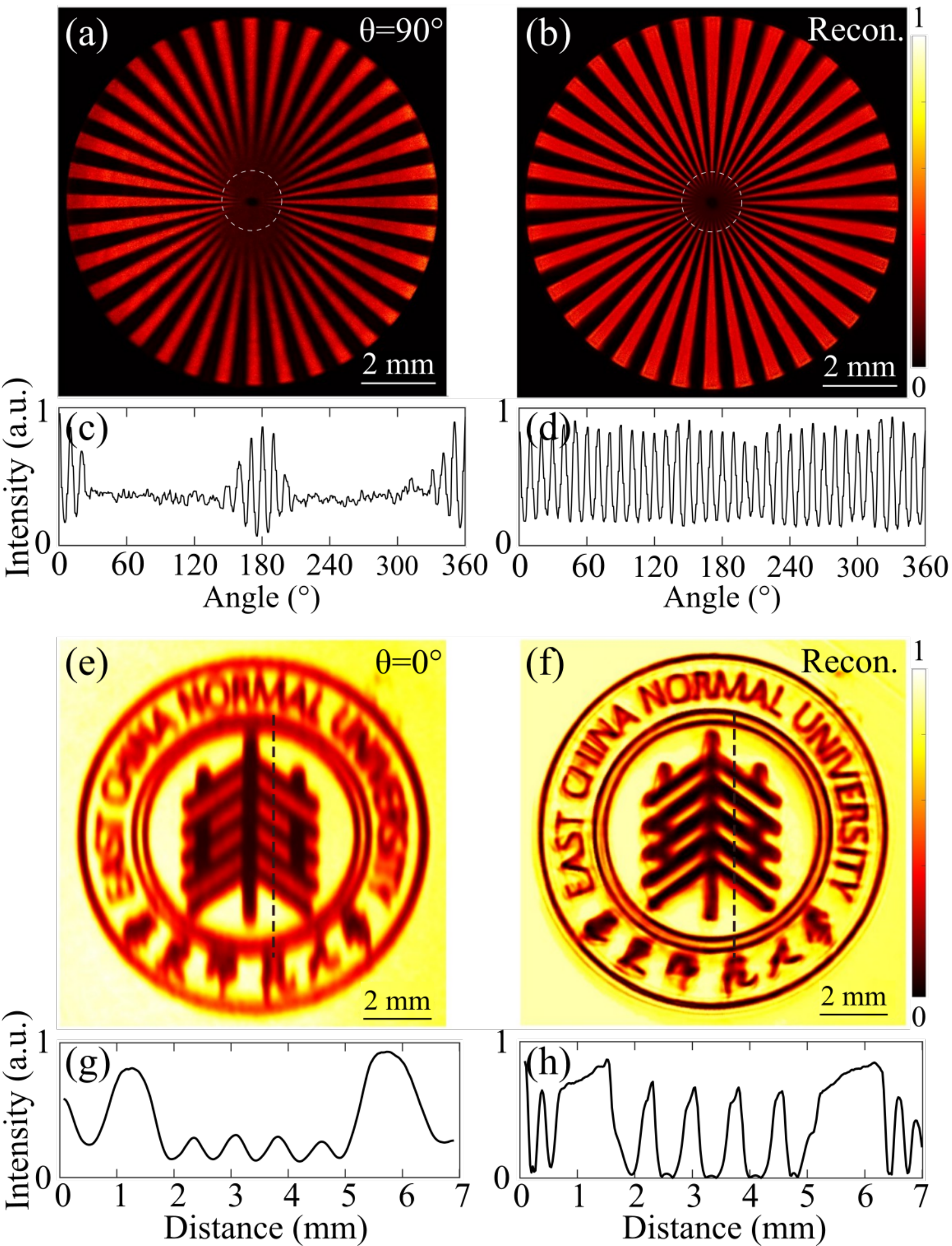}
\caption{Single-photon MIR Fourier ptychographic imaging. (a) Recorded upconversion imaging for the star target under an elliptical pumping at $\theta$=90$^\circ$ in the case of low-light-level illumination flux with 1 photon/pulse/pixel. (b) Reconstructed high-resolution image. (c, d) Corresponding cross sections along the circle at a radius of 0.6 mm, as denoted in the images (a, b). (e) Measured upconversion image for an etched pattern on a silicon wafer under an elliptical pumping at $\theta$=0$^\circ$. (f) Reconstructed high-resolution image for the university emblem. (g, h) Cross sections along two vertical dashed lines denoted in the images (e, f), respectively.}
\label{fig5}
\end{figure}

\subsection{Single-photon MIR Fourier ptychography}
Finally, we proceed to investigating the detection sensitivity of the implemented MIR upconversion imager. To this end, the MIR signal pulse is attenuated to 1 photon/pulse/pixel by using a series of calibrated neutral density filters, before being illuminated onto the targeted object. Meanwhile, the EMCCD is thermoelectrically cooled to suppress the dark noise. The dark current of the EMCCD is specified to be about 10$^{-4}$ electrons/pixel/second at a high gain of 1000. The superior detection sensitivity is essential to implement the MIR upconversion imaging at the single-photon level \cite{Dam2012NP, Wang2021LPR}. Figure \ref{fig5}(a) presents the upconversion image of a Siemens star target with an integration time of 1.5 s in the case of an elliptical pump at $\theta = 90^\circ$. Through the aperture synthetic operation, the reconstructed image is given in Fig. \ref{fig5}(b) under the photon-starved illumination. From the cross sections in Figs. \ref{fig5}(c) and (d), the high-resolution performance of the single-photon MIR Fourier ptychography is obtained within the whole field of view. The demonstrated superior imaging sensitivity benefits from the low-noise nonlinear conversion and high-sensitivity photon counting. Notably, the nonlinear coincidence pumping favors a high conversion efficiency due to the intensive peak power of the pulsed pump, and facilitates a significant suppression of the background noises within an ultrashort time window \cite{Huang2022NC, Fang2023LSA}.

In comparison to visible or near-infrared wavelengths, one unique feature for the MIR imaging is the ability to penetrate through semiconductor materials within the transparency window \cite{Israelsen2019LSA, Fang2023LSA}. For instance, the silicon and germanium substrates  widely used for chips are opaque for the optical fields at wavelengths below 1.1 and 2.0 $\mu$m, respectively. As a proof-of-principle demonstration, we show the penetration imaging capability through a silicon wafer. The wafer has a thickness of 700 $\mu$m and is polished at the two surfaces. Particularly, a university emblem is inscribed onto the backside with the help of the laser ablation. The stronger scattering at the ablated part will lead to a reduced power transmission, which can generate the object pattern through the transmitted infrared light. Figure \ref{fig5}(e) gives a recorded pattern under an elliptical pumping at $\theta = 0^\circ$. The aperture synthesized image is illustrated in Fig. \ref{fig5}(f). The resolution enhancement is clearly revealed from the cross sections in Figs. \ref{fig5}(g) and (h). The high-sensitivity and high-throughput MIR imaging would provide an effective way to perform non-destructive defect inspection for semiconductors and polymer materials in extreme scenarios for industrial quality control and process monitoring \cite{Israelsen2019LSA, Huang2022NC}.

\begin{table*}[t!]
	\centering
	\caption{Performance comparison of wide-field MIR upconversion imaging systems.}
	\centering
	\renewcommand\arraystretch{1.80}
	\label{tab1}
	\renewcommand\arraystretch{1.80}
	\tabcolsep=0.3cm
	\begin{threeparttable}
		
     \begin{tabular}{c|c|c|c|c|c|c}
	\hline
	$\multirow{2}{*}{Ref.} $ & $\multirow{2}{*}{Crystal}$ & $\multirow{2}{*}{Wavelength}$ & $\multirow{2}{*}{Res.}$ & $\multirow{2}{*}{FOV}$ & \multirow{2}{*}{SBP} & \multirow{2}{*}{Sensitivity} \\
	& &($\mu$m)  & ($\mu$m) & (mm) & &\\
	\hline 
	This & CPLN & 3.07 & 39 & 25 & 3.2$\times$10$^{5}$ & 1 photon/pulse/pixel \\  
	\cite{Huang2022NC} & CPLN & 3.07 & 125 & 24.8 & 3.1$\times$10$^{4}$ & 1 photon/pulse \\ 
	\cite{Ge2023PRAppl} & CPLN & 4.15 & 315 & 29.5 & 6.9$\times$10$^{3}$ & 2.5$\times$10$^{3}$ photons/second/pixel \\ 
	\cite{Junaid2019Optica} & LN & 3-4 & 35 & 10 & 6.4$\times$10$^{4}$ & - \\ 
	\cite{Huot2019AO} & LN & 2-4.2 & 75 & 13 & 2.4$\times$10$^{4}$ & - \\ 
	\cite{Zeng2023LPR} & $\beta$-BBO & 2-3 & 31 & 4.5 & 2.1$\times$10$^{4}$ & 35 photons/pulse/pixel \\ 
	\cite{Kviatkovsky2020SA} & PPKTP & 3.4-4.3 & 320 & 9.2 & 6.5$\times$10$^{2}$ & 2.1$\times$10$^{4}$ photons/second/pixel \\ 
	\hline
    \end{tabular}
			
	\end{threeparttable}
\end{table*}

\section{Discussions and conclusion}
Over the past decades, the upconversion technique has rapidly developed as a promising approach to realize sensitive MIR sensing and imaging at room temperature \cite{Barh2019AOP}. Particularly, the superior detection sensitivity at the single-photon level is made possible due to the emerging QPM technique with superior conversion efficiencies \cite{Dam2012NP}. However, the involved periodical poling nonlinear crystals have long suffered from the technical difficulty to achieve large apertures, especially for preparing inverting structures with short domain periods \cite{Missey1998OL, Ishizuki2012OE, Ishizuki2005OL}. Consequently, the field of view or the spatial resolution in the upconversion imaging system is severely limited by the available aperture size, thus posing a challenge to access a high-throughput imaging performance. To date, one remaining quest for the upconversion imagers lies in improving the spatial resolution over a large field of view, which is highly demanded to promote broader applications \cite{Tamamitsu2024NP, Hiramatsu2019SA, Pavlovetc2020PCCP, Park2021AP}. 

Here, we have addressed this quest to implement a high-performance MIR upconversion imaging based on the Fourier ptychography, which offers desirable features of single-photon sensitivity, high spatial resolution, and wide field of view. In the proposed methodology, we extend the aperture synthetic operation to the elliptical pumping upconversion imaging. The acquired sequence of upconversion images as rotating the elliptical passband allows us to significantly boost the resolving power of the upconversion imager through a computational reconstruction process. For the sake of direct comparison, Table \ref{tab1} presents the imaging performances for various representative systems on the wide-field MIR upconversion imaging. Our work demonstrates resolvable elements as many as 3.2$\times$10$^5$, which indicates an over ten-fold improvement than reported values. Notably, a chirped poling design for the nonlinear crystal is adopted to realize a wide-field imaging under a narrow-band illumination \cite{Huang2022NC}, which contrasts to previous schemes by using large-aperture bulky crystals without poling structures. In those works, the wide-field performance is realized by resorting to nonlinear crystal rotation \cite{Junaid2019Optica}, broadband supercontinuum illumination \cite{Huot2019AO}, or shortened crystal lengths \cite{Zeng2023LPR}. Moreover, the detection sensitivity is limited by the relatively small conversion efficiency at the birefringent phase-matching condition.

To go beyond the achieved performance, there are several feasible improvements to be investigated in the future. First, the achieved spatial resolution is currently limited by the numerical aperture of the imaging lenses. The use of microscope objectives with an optimized design for correcting distortion aberrations would make it possible to access the refraction-limited resolution. Benefiting from advanced fabrication technology of lithium niobate wafers, the crystal width of the nonlinear crystal can readily reach to above 20 mm \cite{Nelson1996IEEE}. The resultant size of the synthetic aperture imposes a negligible effect on the image blurring effect. Second, the ability of Fourier ptychography to increase the SBP comes with the price of a reduced temporal imaging resolution, since the technique must acquire multiple images of the sample. The full coverage of targeted Fourier space is essential to reconstruct high-quality images (see discussions in Supplement 1, Note 3). Actually, the overlap of rotated elliptical apertures during adjacent measurements is the key to ensure a stable and fast image reconstruction, which both avoids ambiguities in the solution and accelerates the convergence speed of the phase retrieval process \cite{Zheng2021NRP}. The involved measurement overhead due to the data redundancy requirement can be alleviated by resorting to more sophisticated sampling strategies \cite{Sun2016OE}. Third, the operation wavelength for the CPLN crystal lies in the transparent window up to 5 $\mu$m (see Supplement 1, Note 5), which can be extended to the far-infrared region beyond 13 $\mu$m, by using QPM crystals based on orientation-patterned gallium arsenide (OP-GaAs) or gallium phosphide (OP-GaP) \cite{Demur2017OL}. In the presence of tunable \cite{Junaid2019Optica} or broadband \cite{Fang2024NC, Zhao2023NC} infrared illumination sources, the implemented setup can be immediately extended to perform the hyperspectral imaging for acquiring morphological and chemical information of the samples \cite{Cheng350Science, Shi2020NM}. Notably, the radial dispersion effect in the 4f upconversion imaging configuration would result in a wavelength-dependent spatial magnification, which can be numerically compensated with the pre-determined scaling factor \cite{Fang2024NC}. A preliminary demonstration of high-resolution MIR imaging for a biological sample is presented by examining the venation structure on a dragonfly wing, as given in Supplement 1, Note 6.
 
Moreover, the presented upconversion imaging system can be readily modified to include more functional modalities. In principle, the Fourier ptychographic operation allows full recovery of the complex optical field that contains both the intensity and phase information of the sample \cite{Konda2020OE, Zheng2021NRP}. Therefore, the ptychography-based upconversion imaging not only enhances the spatial resolution across a large field of view, but also provides an effective way to retrieve the phase information of the infrared target. Additionally, the picosecond pump pulse serves as an ultrafast optical gate of the upconversion imager, which can facilitate the depth-resolved imaging by identifying incident photons with different time delays \cite{Fang2023LSA}. Therefore, a novel paradigm of high-dimensional MIR imaging could be foreseen to offer high detection sensitivity, wide field of view and high spatial resolution, which would be desirable for high-throughput applications in material, chemical, and biomedical fields \cite{Tian2015Optica}.\\

\vspace{16pt}
\noindent  {\fontfamily{phv}\selectfont 
\normalsize \textbf{Funding.} 
}
\noindent {\normalsize Shanghai Pilot Program for Basic Research (TQ20220104); National Key R\&D Program (2021YFB2801100), National Natural Science Foundation of China (62175064, 62235019, 62035005); Natural Science Foundation of Chongqing (CSTB2023NSCQ-JQX0011, CSTB2022TIAD-DEX0036); Shanghai Municipal Science and Technology Major Project (2019SHZDZX01); Fundamental Research Funds for the Central Universities.}

\vspace{8pt}
\noindent  {\fontfamily{phv}\selectfont 
\normalsize \textbf{Disclosures.} 
}
\noindent {\normalsize The authors declare no conflicts of interest.}

\vspace{8pt}
\noindent  {\fontfamily{phv}\selectfont 
\normalsize \textbf{Data availability.} 
}
\noindent {\normalsize Data underlying the results presented in this paper are not publicly available at this time but may be obtained from the authors upon reasonable request.}

\vspace{8pt}
\noindent  {\fontfamily{phv}\selectfont 
\normalsize \textbf{Supplemental document.} 
}
\noindent {\normalsize See Supplement 1 for supporting content.}



\begin{thebibliography}{100}

\bibitem{Vodopyanov2020Book} K. L. Vodopyanov, ``Laser-based mid-infrared sources and applications," \textit{John Wiley $\&$ Sons}, (2020).

\bibitem{Cheng350Science} J.-X. Cheng and X. S. Xie, ``Vibrational spectroscopic imaging of living systems: An emerging platform for biology and medicine," \textit{Science} \textbf{350}, aaa8870 (2015).

\bibitem{Hermes2018JO} M. Hermes, R. B. Morrish, L. Huot, L. Meng, S. Junaid, J. Tomko, G. R. Lloyd, W. T. Masselink, P. Tidemand-Lichtenberg, C. Pedersen, F. Palombo, and N. Stone, ``Mid-IR hyperspectral imaging for label-free histopathology and cytology," \textit{J. Opt.} \textbf{20}, 023002 (2018).

\bibitem{Lau2023CSR} J. A. Lau, V. B. Verma, D. Schwarzer, and A. M. Wodtke,  ``Superconducting single-photon detectors in the mid-infrared for physical chemistry and spectroscopy," \textit{Chem. Soc. Rev.} \textbf{52}, 921 (2023).

\bibitem{Hadfield2023Optica} R. H. Hadfield, J. Leach, F. Fleming, D. J. Paul, C. H. Tan, J. S. Ng, R. K. Henderson, and G. S. Buller. ``Single-photon detection for long-range imaging and sensing," \textit{Optica} \textbf{10}, 1124 (2023).

\bibitem{Israelsen2019LSA} N. M. Israelsen, C. R. Petersen, A. Barh, D. Jain, M. Jensen, G. Hannesschl$\ddot{a}$ger, P. Tidemand-Lichtenberg, C. Pedersen, A. Podoleanu, and O. Bang, ``Real-time high-resolution mid-infrared optical coherence tomography," \textit{Light Sci. Appl.} \textbf{8}, 11 (2019).

\bibitem{Shi2020NM} L. Shi, X. Liu, L. Shi, H. T.  Stinson, J. Rowlette, L. J. Kahl, C. R. Evans, C. Zheng, L. E. P. Dietrich, and W. Min, ``Mid-infrared metabolic imaging with vibrational probes," \textit{Nat. Methods} \textbf{17}, 844 (2020).

\bibitem{Wolley2023PNAS} O. Wolley, S. Mekhail, P.-A. Moreau, T. Gregory, G. Gibson, G. Leuchs, and M. J. Padgett, ``Near single-photon imaging in the shortwave infrared using homodyne detection," \textit{Proc. Natl Acad. Sci. USA} \textbf{120}, e2216678120 (2023).    
  
\bibitem{Razeghi2014RPP} M. Razeghi and B. M. Nguyen, ``Advances in mid-infrared detection and imaging: a key issues review," \textit{Rep. Prog. Phys.} \textbf{77}, 082401 (2014).

\bibitem{Wang2019Small} P. Wang, H. Xia, Q. Li, F. Wang, L. Zhang, T. Li, P. Martyniuk, A. Rogalski, and W. Hu,  ``Sensing infrared photons at room temperature: from bulk materials to atomic layers," \textit{Small}, \textbf{15}, 46 (2019).


\bibitem{Wu2021NR} J. Wu, N. Wang, X. Yan, and H. Wang,  ``Emerging low-dimensional materials for mid-infrared detection," \textit{Nano Res.} \textbf{14}, 1863 (2021).

\bibitem{Taylor2023Optica} G. G. Taylor, A. B. Walter, B. Korzh, B. Bumble, S. R. Patel, J. P. Allmaras, A. D. Beyer, R. O$^{'}$Brient, M. D. Shaw, and E. E. Wollman, ``Low-noise single-photon counting superconducting nanowire detectors at infrared wavelengths up to 29 $\mu$m," \textit{Optica} \textbf{10}, 1672 (2023).

\bibitem{Chen2021SB} Q. Chen, R. Ge, L. Zhang, F. Li, B. Zhang, F. Jin, H. Han, Y. Dai, G. He, Y. Fei, X. Wang, H. Wang, X. Jia, Q. Zhao, X. Tu, L. Kang, J. Chen, and P. Wu, ``Mid-infrared single photon detector with superconductor Mo$_{80}$Si$_{20}$ nanowire," \textit{Sci. Bull.} \textbf{66}, 965 (2021). 
 
\bibitem{Pan2022OE} Y. Pan, H. Zhou, X. Zhang, H. Yu, L. Zhang, M. Si, H. Li, L. You, and Z. Wang, ``Mid-infrared Nb$_{4}$N$_{3}$-based superconducting nanowire single photon detectors for wavelengths up to 10 $\mu$m," \textit{Opt. Express} \textbf{30}, 40044 (2022).
  
\bibitem{Hampel2024APL} B. Hampel, R. P. Mirin, S. W. Nam, and V.B. Verma, ``A 64-pixel mid-infrared single-photon imager based on superconducting nanowire detectors,"  \textit{Appl. Phys. Lett.} \textbf{124}, 042602 (2024).

\bibitem{Tamamitsu2024NP} M. Tamamitsu, K. Toda, M. Fukushima, V. R. Badarla, H. Shimada, S. Ota, K. Konishi, and T. Ideguchi, ``Mid-infrared wide-field nanoscopy," \textit{Nat. Photonics} \textbf{17}, 738 (2024).

\bibitem{Hiramatsu2019SA} K. Hiramatsu, T. Ideguchi, Y. Yonamine, S. Lee, Y. Luo, K. Hashimoto, T. Ito, M. Hase, J.-W. Park, Y. Kasai, S. Sakuma, T. Hayakawa, F. Arai, Y. Hoshino, and K. Goda, ``High-throughput label-free molecular fingerprinting flow cytometry," \textit{Sci. Adv.} \textbf{5}, eaau0241 (2019). 

\bibitem{Pavlovetc2020PCCP} I. M. Pavlovetc, K. Aleshire, G. V. Hartland, and M. Kuno, ``Approaches to mid-infrared, super-resolution imaging and spectroscopy," \textit{Phys. Chem. Chem. Phys.} \textbf{22}, 4313 (2020).

\bibitem{Park2021AP} J. Park, D. J Brady, G. Zheng, L. Tian, and L. Gao, ``Review of bio-optical imaging systems with a high space-bandwidth product," \textit{Adv. Photonics} \textbf{3}, 044001 (2021).

\bibitem{Barh2019AOP} A. Barh, P. J. Rodrigo, L. Meng, C. Pedersen, and P. Tidemand-Lichtenberg, ``Parametric upconversion imaging and its applications," \textit{Adv. Opt. Photonics} \textbf{11}, 952 (2019).


\bibitem{Dam2012NP} J. S. Dam, P. Tidemand-Lichtenberg, and C. Pedersen,  ``Room-temperature mid-infrared single-photon spectral imaging," \textit{Nat. Photonics} \textbf{6}, 788 (2012).

\bibitem{Wang2021LPR} Y. Wang, J. Fang, T. Zheng, Y. Liang, Q. Hao, E. Wu, M. Yan, K. Huang, and H. Zeng, ``Mid-Infrared Single-Photon Edge Enhanced Imaging Based on Nonlinear Vortex Filtering," \textit{Laser Photon. Rev.} \textbf{15}, 2100189 (2021).

\bibitem{Mrejen2020LPR} M. Mrejen, Y. Erlich, A. Levanon, and H. Suchowski,  ``Multicolor Time-Resolved Upconversion Imaging by Adiabatic Sum Frequency Conversion," \textit{Laser Photon. Rev.} \textbf{14}, 2000040 (2020).

\bibitem{Huot2019AO} L. Huot, P. M. Moselund, P. Tidemand-Lichtenberg, and C. Pedersen, ``Pulsed upconversion imaging of mid-infrared supercontinuum light using an electronically synchronized pump laser," \textit{Appl. Opt.} \textbf{58}, 244 (2019).

\bibitem{Zeng2023LPR} X. Zeng, C. Wang, H. Wang, Q. Lin, Z. Chen, X. Lu, M. Zheng, J. Liang, Y. Cai, S. Xu, and J. Li, ``Tunable Mid-Infrared Detail-Enhanced Imaging With Micron-Level Spatial Resolution and Photon-Number Resolving Sensitivity," \textit{Laser Photon. Rev.} \textbf{17}, 2200446 (2023).

\bibitem{Paterova2020SA} A. V. Paterova, S. M. Maniam, H. Yang, G. Grenci, and L. A. Krivitsky, ``Hyperspectral infrared microscopy with visible light," \textit{Sci. Adv.} \textbf{6}, eabd0460 (2020).

\bibitem{Kviatkovsky2020SA} I. Kviatkovsky, H. M. Chrzanowski, E. G. Avery, H. Bartolomaeus, and S. Ramelow,  ``Microscopy with undetected photons in the mid-infrared," \textit{Sci. Adv.} \textbf{6}, eabd0264 (2020).

\bibitem{Rehain2020NC} P. Rehain, Y. M. Sua, S. Zhu, I. Dickson, B. Muthuswamy, J. Ramanathan, A. Shahverdi, and Y.-P. Huang, ``Noise-tolerant single photon sensitive three-dimensional imager," \textit{Nat. Commun.} \textbf{11}, 921 (2020).

\bibitem{Ge2023PRAppl} Z. Ge, Z. Han, Y. Liu, X. Wang, Z. Zhou, F. Yang, Y. Li, Y. Li, L. Chen, W. Li, S. Niu, and B. Shi, ``Midinfrared up-conversion imaging under different illumination conditions," \textit{Phys. Rev. Appl.} \textbf{20}, 054060 (2023).

\bibitem{Huang2022NC} K. Huang, J. Fang, M. Yan, E. Wu, and H. Zeng,  ``Wide-field mid-infrared single-photon upconversion imaging," \textit{Nat. Commun.} \textbf{13}, 1077 (2022).


\bibitem{Fang2023LSA} J. Fang, K. Huang, E. Wu,  M. Yan, and  H. Zeng, ``Mid-infrared single-photon 3D imaging," \textit{Light Sci. Appl.} \textbf{12}, 144 (2023).

\bibitem{Zhao2023NC} Y. Zhao, S. Kusama, Y. Furutani, W.-H. Huang, C.-W. Luo, and T. Fuji, ``High-speed scanless entire bandwidth mid-infrared chemical imaging," \textit{Nat. Commun.} \textbf{14}, 3929 (2023).

\bibitem{Fang2024NC} J. Fang, K. Huang, R. Qin, Y. Liang, E. Wu, M. Yan, and H. Zeng, ``Wide-field mid-infrared hyperspectral imaging beyond video rate," \textit{Nat. Commun.} \textbf{15}, 1811 (2024).
  
\bibitem{Junaid2019Optica} S. Junaid, S. C. Kumar, M. Mathez, M. Hermes, N. Stone, N. Shepherd, M. Ebrahim-Zadeh, P. Tidemand-Lichtenberg, and C. Pedersen, ``Video-rate, mid-infrared hyperspectral upconversion imaging," \textit{Optica} \textbf{6}, 702 (2019).

\bibitem{Knez2022SA} D. Knez, B. W. Toulson, A. Chen, M. H. Ettenberg, H. Nguyen, E. O. Potma, and D. A. Fishman, ``Spectral imaging at high definition and high speed in the mid-infrared," \textit{Sci. Adv.} \textbf{8}, eade4247 (2022).

\bibitem{Molina2024AM} L. V. Molina, R. C. Morales, J. Zhang, R. Schiek, I. Staude, A. A. Sukhorukov, and D. N. Neshev, ``Enhanced Infrared Vision by Nonlinear Up-Conversion in Nonlocal Metasurfaces," \textit{Adv. Mater.} \textbf{36}, 2402777 (2024).
  
\bibitem{Bahabad2010NP} A. Bahabad, M. M. Murnane, and  H. C. Kapteyn, ``Quasi-phase-matching ofmomentum and energy in nonlinear optical processes," \textit{Nat. Photonics} \textbf{4}, 570 (2010).

\bibitem{Missey1998OL} M. J. Missey, V. Dominic, L. E. Myers, and R. C. Eckardt, ``Diffusion-bonded stacks of periodically poled lithium niobate," \textit{Opt. Lett.} \textbf{23},  664 (1998).

\bibitem{Ishizuki2012OE} H. Ishizuki and T. Taira, ``Half-joule output optical-parametric oscillation by using 10-mm-thick periodically poled Mg-doped congruent LiNbO$_{3}$," \textit{Opt. Express} \textbf{20}, 20002 (2012).
  
\bibitem{Ishizuki2005OL} H. Ishizuki and T. Taira, ``High-energy quasi-phase-matched optical parametric oscillation in a periodically poled MgO:LiNbO$_{3}$ device with a 5mm $\times$ 5mm aperture," \textit{Opt. Lett.} \textbf{30}, 2918 (2005).


\bibitem{Huang2024LSA} Z. Huang and L. Cao, ``Quantitative phase imaging based on holography: trends and new perspectives," \textit{Light Sci. Appl.} \textbf{13}, 145 (2024).
  
\bibitem{Pan2020RPP} A. Pan, C. Zuo, and B. Yao, ``High-resolution and large field-of-view Fourier ptychographic microscopy and its applications in biomedicine," \textit{Rep. Prog. Phys.} \textbf{83}, 096101 (2020).
    
\bibitem{Konda2020OE} P. C. Konda, L. Loetgering, K. C. Zhou, S. Xu, A. R. Harvey, and R. Horstmeyer, ``Fourier ptychography: current applications and future promises," \textit{Opt. Express} \textbf{28}, 9603 (2020).

\bibitem{Zheng2021NRP} G. Zheng, C. Shen, S. Jiang, P. Song, and C. Yang, ``Concept, implementations and applications of Fourier ptychography," \textit{Nat. Rev. Phys.} \textbf{3}, 207 (2021).
  
\bibitem{Zheng2013NP} G. Zheng, R. Horstmeyer, and C. Yang, ``Wide-field, high-resolution Fourier ptychographic microscopy," \textit{Nat. Photonics} \textbf{7}, 739 (2013).

\bibitem{Holloway2017SA} J. Holloway, Y. Wu, M. K. Sharma, O. Cossairt, and A. Veeraraghavan, ``SAVI: Synthetic apertures for long-range, subdiffraction-limited visible imaging using Fourier ptychography," \textit{Sci. Adv.} \textbf{3}, e1602564 (2017).

\bibitem{Nelson1996IEEE} M. D. Nelson, V. Dominic, T. P. Grayson, and L. E. Myers, ``PPLN optical parametric oscillator energy scaling capabilities," \textit{Proceedings of the IEEE 1996 National Aerospace and Electronics Conference}.

\bibitem{Sun2016OE} J. Sun, Q. Chen, Y. Zhang, and C. Zuo, ``Sampling criteria for Fourier ptychographic microscopy in object space and frequency space," \textit{Opt. Express} \textbf{24}, 15765 (2016).

\bibitem{Demur2017OL} R. Demur, A. Grisard, L. Morvan, E. Lallier, N. Treps, and C. Fabre, ``High sensitivity narrowband wavelength mid-infrared detection at room temperature," \textit{Opt. Lett.} \textbf{42}, 2006 (2017).

\bibitem{Tian2015Optica} L. Tian and L. Waller, ``3D intensity and phase imaging from light field mea-surements in an LED array microscope," \textit{Optica} \textbf{2}, 104 (2015).

\end{thebibliography}
\end{document}